\documentclass[12pt]{article}

\begin{document}

\begin{center}
{\LARGE Algebraic spectral relations for elliptic quantum Calogero--Moser
problems\vspace{0.7cm}}
\end{center}

{\large L. A. Khodarinova}$^{{\dag }}${\large \ and\ I. A. Prikhodsky}$^{{%
\ddag }}${\large \ \vspace{0.3cm}}

$^{{\dag }}$\textit{Magnetic Resonance Centre, School of Physics and
Astronomy, University of Nottingham, Nottingham, England NG7 2RD, e-mail:
LarisaKhodarinova@hotmail.com}

$^{{\ddag }}$\textit{Institute of Mechanical Engineering, Russian Academy of
Sciences,M. Haritonievsky, 4, Centre, Moscow 101830 Russia }

\vspace{1.0cm}

\textbf{Abstract.} Explicit algebraic relations between the quantum
integrals of the elliptic Calogero--Moser quantum problems related to the
root systems $\mathbf{A_2}$ and $\mathbf{B_2}$ are found. \vspace{0.3cm}

\section*{1. Introduction.}

The notion of algebraic integrability for the quantum problems arised as a
multidimensional generalization of the "finite--gap" property of the
one-dimensional Schr\"odinger operators (see~\cite{Kr},~\cite{ChV}, ~\cite%
{VSCh}).

Recall that the Schr\"odinger equation 
\[
L\psi=-\Delta\psi+u(x)\psi=E\psi ,\qquad x\in R^n , 
\]
is called \textit{integrable} if there exist $n$ commuting differential
operators $L_1=L,L_2,\ldots,L_n$ with the constant algebraically independent
highest symbols $P_1(\xi)=\xi^2,P_2(\xi),\ldots,P_n(\xi)$, and \textit{%
algebraically integrable} if there exists at least one more differential
operator $L_{n+1}$, which commutes with the operators $L_i, i=1,\ldots,n$,
and whose highest symbol $P_{n+1}(\xi)$ is constant and takes different
values on the solutions of the algebraic system $P_i(\xi)=c_i, i=1,\ldots,n$%
, for generic $c_i$.

According to the general result \cite{Kr} in the algebraically integrable
case there exists an algebraic relation between the operators $L_i,
i=1,\ldots,n+1$:

\[
Q(L_1,L_2,\ldots,L_{n+1}) = 0. 
\]

The corresponding eigenvalues $\lambda_i$ of the operators $L_i$ obviously
satisfy the same relation: 
\[
Q(\lambda_1,\lambda_2,\ldots,\lambda_{n+1}) = 0. 
\]

Sometimes it is more suitable to add more generators from the commutative
ring of quantum integrals: $L_1,L_2,\ldots,L_{n+k}$; in that case we have
more than one relation. We will call these relations \textit{spectral}. They
determine a spectral variety of the corresponding Schr\"odinger operator.

The main result of the present paper is the explicit description of the
spectral algebraic relations for the three--particle elliptic
Calogero--Moser problem with the Hamiltonian 
\begin{equation}
L=-\frac{\partial ^2}{\partial {x_1^2}}-\frac{\partial ^2}{\partial {x_2^2}}-%
\frac{\partial ^2}{\partial {x_3^2}}+4(\wp (x_1-x_2)+\wp (x_2-x_3)+\wp
(x_3-x_1))  \label{1}
\end{equation}
and for the generalised Calogero--Moser problem related to the root system $%
\mathbf{B_2}$ (see~\cite{OP}) with the Hamiltonian 
\begin{equation}
L=-\frac{\partial ^2}{\partial {x_1^2}}-\frac{\partial ^2}{\partial {x_2^2}}%
+2(\wp (x_1)+\wp (x_2)+2\wp (x_1+x_2)+2\wp (x_1-x_2))  \label{2}
\end{equation}
Here $\wp $ is the classical Weierstrass elliptic function satisfying the
equation 
\[
\wp ^{\prime }{}^2-4\wp ^3+g_2\wp +g_3=0. 
\]
These operators are known as the simplest multidimentional generalizations
of the classical Lame operator (see~\cite{ChV}).

For the problem~(\ref{1}) reduced to the plane $x_1 + x_2 + x_3 = 0$ the
explicit equations of the spectral variety have been found before in~\cite%
{SV}. The derivation of ~\cite{SV} is indirect and based on the idea of the
"isoperiodic deformations"~\cite{GS}. In the present paper we give a direct
derivation of the spectral relations for the problem~(\ref{1}) using
explicit formulae for the additional quantum integrals, which have been
found in~\cite{Kh}. The derived formulae are in a good agreement with the
formulae written in~\cite{SV}.

For the elliptic Calogero--Moser problem related to the root system $\mathbf{%
B_2}$ the algebraic integrability with the explicit formulae for the
additional integrals were obtained in the recent paper~\cite{O}. We use
these formulae to derive the spectral relations for the operator~(\ref{2}).

We would like to mention that although the procedure of the derivation of
the spectral relations provided the quantum integrals are given is
effective, the actual calculations are huge and would be very difficult to
perform without computer. We have used a special program, which has been
created for this purpose.

\section*{2. Spectral relations for the three-particle elliptic
Calogero--Moser problem.}

Let's consider the quantum problem with the Hamiltonian~(\ref{1}). This is a
particular case of the three-particle elliptic Calogero--Moser problem
corresponding to a special value of the parameter in the interaction. The
algebraic integrability in this case has been conjectured by Chalykh and
Veselov in~\cite{ChV} and proved later in~\cite{SV},~\cite{Kh} and (in much
more general case) in~\cite{BEG}. We should mention that only the paper~\cite%
{Kh} contains the explicit formulae for the additional integrals, the other
proofs are indirect.

The usual integrability of this problem has been established by Calogero,
Marchioro and Ragnisco in~\cite{CMR}. The corresponding integrals have the
form 
\[
\begin{array}{l}
L_1=L=-\Delta +4(\wp _{12}+\wp _{23}+\wp _{31}), \\ 
L_2=\partial _1+\partial _2+\partial _3, \\ 
L_3=\partial _1\partial _2\partial _3+2\wp _{12}\partial _3+2\wp
_{23}\partial _1+2\wp _{31}\partial _2,%
\end{array}
\]
where we have used the notations $\partial _i={\partial }/{\partial x_i}$, $%
\wp _{ij}=\wp (x_i-x_j)$.

The following additional integrals have been found in~\cite{Kh}: 
\[
\begin{array}{l}
I_{12}=(\partial _1-\partial _3)^2(\partial _2-\partial _3)^2-8\wp
_{23}(\partial _1-\partial _3)^2-8\wp _{13}(\partial _2-\partial _3)^2 \\ 
\qquad {}+4(\wp _{12}-\wp _{13}-\wp _{23})(\partial _1-\partial _3)(\partial
_2-\partial _3) \\ 
\qquad {}-2(\wp _{12}^{\prime }+\wp _{13}^{\prime }+6\wp _{23}^{\prime
})(\partial _1-\partial _3) \\ 
\qquad {}-2(-\wp _{12}^{\prime }+6\wp _{13}^{\prime }+\wp _{23}^{\prime
})(\partial _2-\partial _3) \\ 
\qquad {}-2\wp _{12}^{\prime \prime }-6\wp _{13}^{\prime \prime }-6\wp
_{23}^{\prime \prime }+4(\wp _{12}^2+\wp _{13}^2+\wp _{23}^2) \\ 
\qquad {}+8(\wp _{12}\wp _{13}+\wp _{12}\wp _{23}+7\wp _{13}\wp _{23}),%
\end{array}
\]
two other integrals $I_{23}$, $I_{31}$ can be written simply by permuting
the indices. Unfortunately, none of these operators implies algebraic
integrability, because the symbols do not take different values on the
solutions of the system $P_i(\xi )=c_i,i=1,2,3$ (see the definition in the
Introduction). But any non-symmetric linear combination of them, e.g. $%
L_4=I_{12}+2I_{23}$ would fit into the definition.

\textbf{Lemma. } \textit{The operators} $L_1,L_2,L_3,I$, \textit{where} $I$ 
\textit{is equal to} $I_{12},I_{23}$ \textit{or} $I_{13}$, \textit{satisfy
the algebraic relation:} 
\begin{equation}  \label{3}
\begin{array}{l}
Q(L_1,L_2,L_3,I) = I^3 + A_1I^2 + A_2I + A_3 = 0,%
\end{array}%
\end{equation}
\[
\begin{array}{l}
\!\!\!\!\!\mathit{where}\; A_1 = 6g_2-X^2, A_2 = 2XY-15g_2^2-2g_2X^2, \\ 
\qquad{} A_3 = -Y^2-2g_2XY-108g_3Y+16g_3X^3+15g_2^2X^2-100g_2^3, \\ 
\qquad{} X = 3/2L_1 + 1/2L_2^2, \\ 
\qquad{} Y = 1/2L_1^3 + 27L_3^2 + 1/4L_2^6 + L_1L_2^4 - 5L_2^3L_3 +
5/4L_1^2L_2^2 - 9L_1L_2L_3.%
\end{array}
\]

The idea of the proof is the following. Let the relation $Q$ be a polynomial
of third degree of $I$ such that $I_{12},I_{23},I_{13}$ are its roots: 
\[
Q = I^3 + A_1I^2 + A_2I + A_3. 
\]
Then $A_1 = -(I_{12}+I_{23}+I_{13})$, $A_2 = I_{12}I_{23} + I_{12}I_{13} +
I_{23}I_{13}$, $A_3 = -I_{12}I_{23}I_{13}$. From the explicit formulae for $%
I_{ij}$ it follows that the operators $A_i$ are symmetric and their highest
symbols $a_i$ are constants. So there exist polynomials $p_i$ such that $a_i
= p_i(l_1,l_2,l_3)$. Consider $A^{\prime}_i = A_i - p_i(L_1,L_2,L_3)$, $%
A^{\prime}_i$ commute with $L$. It follows from the Berezin's lemma~\cite{B}
that if a differential operator commutes with a Schr\"odinger operator then
the coefficients in the highest symbol of this operator are polynomials in $%
x $. Since the coefficients in the highest symbols of the operators $%
A^{\prime}_i$ are some elliptic functions, they must be constant in $x$. It
is clear that the operators $A^{\prime}_i$ are also symmetric and $\deg
a^{\prime}_i < \deg a_i$. So we can continue this procedure until we come to
zero. Thus we express $A_i$ as the polynomials of $L_1$, $L_2$ and $L_3$. To
calculate the explicit expressions we can use the fact that the coefficients
in the highest symbols $a^{\prime}_i$ are constant and therefore may be
calculated at some special point. The most suitable choice is when $%
x_1-x_2=\omega_1, x_2-x_3=\omega_2, x_1-x_3=(\omega_1+\omega_2) $,where $%
2\omega_1, 2\omega_2$ are the periods of Weierstrass $\wp$-function. Then $%
\wp^{\prime}(x_1-x_2)=\wp^{\prime}(x_2-x_3)=\wp^{\prime}(x_1-x_3)=0$ and $%
\wp(x_1-x_2)=e_1,\wp(x_2-x_3)=e_2,\wp(x_1-x_3)=e_3,$ where $e_1,e_2,e_3$ are
the roots of the polynomial $4z^3 - g_2 z - g_3 = 0$ (see e.g.~\cite{WW})

\textbf{Theorem 1. } \textit{The operators} $L_1, L_2, L_3$, $I = I_{12}$, $%
J = I_{23}$, $L_4 = I+2J$ \textit{satisfy the algebraic system:} 
\[
\begin{array}{l}
I^3 + A_1I^2 + A_2I + A_3 = 0, \\ 
I^2 + IJ + J^2 + A_1(I+J) + A_2 = 0,%
\end{array}
\]
\textit{where} $A_1, A_2, A_3$ \textit{are given by the formulae}~(\ref{3}).

\textbf{Proof.} The first relation for $I$ is proved in the lemma. The proof
of the second relation is the following. The operators $I_{12} = I$, $I_{23}
= J $, $I_{13}$ satisfy the relations: $I+J+I_{13} = -A_1$, $IJ +
(I+J)I_{13} = A_2$, so $I_{13} = -(I+J+A_1)$ and therefore we obtain $IJ -
(I+J)(I+J+A_1) = A_2$. This completes the proof of Theorem 1.

\textit{Remark.} Putting $L_2=0$ we can reduce the problem~(\ref{1}) to the
plane $x_1+x_2+x_3=0$ and obtain two-dimensional elliptic Calogero--Moser
problem related to the root system $\mathbf{A_2}$. The spectral curve of
this problem was found in the paper~\cite{SV}. The corresponding formula
from~\cite{SV} has the form: 
\[
\nu^3 + (6\lambda\mu^2-3(\lambda^2-3g_2)^2)\nu
-\mu^4+(10\lambda^3-18g_2\lambda+108g_3)\mu^2 + 2(\lambda^2-3g_2)^3 = 0 
\]
If in our formula~(\ref{3}) put $L_2=0$, substitute $L_1=2\lambda$, $L_3=%
\sqrt{3}/9\mu$, $I=\nu-1/3(6g_2-X^2)$, then for the relation~(\ref{3}) we
get: 
\[
\nu^3 + (6\lambda\mu^2-3(\lambda^2-3g_2)^2)\nu
-\mu^4+(10\lambda^3-18g_2\lambda-108g_3)\mu^2 + 2(\lambda^2-3g_2)^3 = 0 
\]
The elliptic curves $y^2 = 4x^3-g_2x-g_3$ and $y^2 = 4x^3-g_2x+g_3$ are
isomorphic: $x\to -x, y\to iy$, so the difference in the sign by $g_3$ is
not important.

\section*{3. Spectral relations for the elliptic Calogero--Moser problem
related to the root system $\mathbf{B_{2}}$.}

Consider now the Schr\"odinger operator~(\ref{2}). Its algebraic
integrability conjectured in~\cite{ChV} has been proved in~\cite{O}.

The formulae for the quantum integrals in this case are (see~\cite{O}) 
\[
\begin{array}{l}
L_1=L=-\Delta +2(\wp (x)+\wp (y)+2\wp (x+y)+2\wp (x-y)), \\ 
L_2=\partial _x^2\partial _y^2-2\wp (y)\partial _x^2-2\wp (x)\partial
_y^2-4(\wp (x+y)-\wp (x-y))\partial _x\partial _y \\ 
\qquad {}-2(\wp ^{\prime }(x+y)+\wp ^{\prime }(x-y))\partial _x-2(\wp
^{\prime }(x+y)-\wp ^{\prime }(x-y))\partial _y \\ 
\qquad {}-2(\wp ^{\prime \prime }(x+y)+\wp ^{\prime \prime }(x-y))+4(\wp
^2(x+y)+\wp ^2(x-y)) \\ 
\qquad {}+4(\wp (x)+\wp (y))(\wp (x+y)+\wp (x-y)) \\ 
\qquad {}-8\wp (x+y)\wp (x-y)-4\wp (x)\wp (y), \\ 
L_3=I_x+2I_y,\;\;\mathrm{where}%
\end{array}
\]
\[
\begin{array}{l}
I_x=\partial _x^5-5\partial _x^3\partial _y^2-10(1/2\wp (x)-\wp (y)+\wp
(x+y)+\wp (x-y))\partial _x^3 \\ 
\qquad {}+30(\wp (x+y)-\wp (x-y))\partial _x^2\partial _y+15\wp (x)\partial
_x\partial _y^2 \\ 
\qquad {}-15/2\wp (x)(\partial _x^2-\partial _y^2)+30(\wp (x+y)-\wp
(x-y))\partial _x\partial _y \\ 
\qquad {}+(10\wp ^{\prime \prime }(x+y)-10\wp ^{\prime \prime }(x-y)-30\wp
(y)(\wp (x+y)-\wp (x-y)))\partial _y \\ 
\qquad {}+(30\wp (y)(\wp (x)-\wp (x+y)-\wp (x-y))+120\wp (x+y)\wp (x-y) \\ 
\qquad {}+10\wp ^{\prime \prime }(x+y)+10\wp ^{\prime \prime }(x-y)-5\wp
^{\prime \prime }(x)-9/2g_2)\partial _x \\ 
\qquad {}-15(\wp ^{\prime }(x+y)+\wp ^{\prime }(x-y))(\wp (x)+\wp (y)) \\ 
\qquad {}-15(\wp ^{\prime }(x)\wp (y)+\wp ^{\prime }(y)\wp (x)) \\ 
\qquad {}+60(\wp ^{\prime }(x+y)+\wp ^{\prime }(x-y))(\wp (x+y)+\wp (x-y)),%
\end{array}
\]
operator $I_y$ can be written by exchanging in the previous formula $x$ and $%
y$, and we use the notations $\partial _x={\partial }/{\partial x},\partial
_y={\partial }/{\partial y}$.

\textbf{Theorem 2. } \textit{The quantum integrals} $L= 1/2 L_1,M= L_2, I =
I_x, J = I_y, L_3 = I+2J$ \textit{of the elliptic Calogero--Moser system
related to the root system} $\mathbf{B_2}$ \textit{satisfy the following
algebraic relations:}

\[
\begin{array}{l}
\qquad{} I^4+B_1I^2+B_2=0, \\ 
\qquad{} I^2+J^2+B_1 = 0, \\ 
\!\!\!\!\!\mathit{where}\; B_1= 32L^5-120ML^3+120M^2L+g_2(-82L^3+114LM) \\ 
\qquad{}\qquad{}+g_3(-270L^2+486M)+102g_2^2L+486g_3g_2, \\ 
\qquad{} B_2=400M^3L^4-1440M^4L^2+1296M^5 \\ 
\qquad{}\qquad{}+g_2(-400ML^6+840M^2L^4 -576M^3L^2+648M^4) \\ 
\qquad{}\qquad{}+g_3(800L^7-8280ML^5+22032M^2L^3 -17496LM^3) \\ 
\qquad{}\qquad{}+g_2^2(800L^6-1815ML^4+3510M^2L^2-3807M^3) \\ 
\qquad{}\qquad{}+g_3g_2(3870L^5+324ML^3-13122M^2L) \\ 
\qquad{}\qquad{}+g_3^2(18225L^4-65610ML^2+59049M^2) \\ 
\qquad{}\qquad{}+g_2^3(-2930L^4+5418ML^2-4536M^2) \\ 
\qquad{}\qquad{}+g_3g_2^2(-21708L^3+26244LM) \\ 
\qquad{}\qquad{}+g_3^3g_2(-65610L^2+118098M)+g_2^4(2772L^2-1539M) \\ 
\qquad{}\qquad{}+21870g_3g_2^3L+59049g_2^2g_3^2-162g_2^5.%
\end{array}
\]

The proof is analogous to the previous case. The first relation is a
polymonial of second order in $I^2$ such that $I_x^2$ and $I_y^2$ are its
roots. Calculations of the coefficients as in the previous case can be done
in the special point $(x,y):$ $x = \omega_1, y = \omega_2$. Then $%
\wp(x)=e_1,\wp(y)=e_2,\wp(x+y)=\wp(x-y)=e_3,$ $\wp^{\prime}(x)=\wp^{%
\prime}(y)=\wp^{\prime}(x-y)=\wp^{\prime}(x-y)=0$.

As we have mentioned already these calculations have been done with the help
of the special computer program created by the authors.

The authors are grateful to A.P.Veselov and O.A.Chalykh for useful
discussions.


\begin{thebibliography}{99}
\bibitem{Kr} {\footnotesize I.M.Krichever \textquotedblright Methods of
algebraic geometry in the theory of nonlinear equations\textquotedblright ,
Uspekhi Mat. Nauk 32(6), 198--245 (1977).}

\bibitem{ChV} {\footnotesize O.A.Chalykh, A.P.Veselov. \textquotedblright
Commutative rings of partial differential operators and Lie
algebras\textquotedblright , Commun. Math. Phys.126, 597--611 (1990).}

\bibitem{VSCh} {\footnotesize A.P.Veselov, K.L.Styrkas, O.A.Chalykh
\textquotedblright Algebraic integrability for Schrodinger equations and
finite reflection groups\textquotedblright , Theor. Math. Phys.94(2),
253--275 (1993).}

\bibitem{OP} {\footnotesize M.A.Olshanetsky, A.M.Perelomov
\textquotedblright Quantum integrables systems related to Lie
algebras\textquotedblright , Phys.Reports 94(6), 313-404 (1983).}

\bibitem{SV} {\footnotesize M.U.Schmidt, A.P.Veselov \textquotedblright
Quantum elliptic Calodgero--Moser problem and deformations of algebraic
surfaces\textquotedblright , Preprint of Freie Universitat, Berlin, (1996).}

\bibitem{GS} {\footnotesize P.Grinevich, M.Schmidt \textquotedblright Period
preserving flows and the module space of periodic solutions of soliton
equations\textquotedblright , Physica D 87, 73-98 (1995).}

\bibitem{Kh} {\footnotesize L.A.Khodarinova \textquotedblright On quantum
elliptic Calogero--Moser problem\textquotedblright , Vestnik Mosc. Univ.,
Ser. Math.and Mech. 53, 5, 16--19 (1998).}

\bibitem{O} {\footnotesize A.A.Oblomkov "Integrability of some quantum
system related to the root system }$B_{2}${\footnotesize ", Vestnik Mosc.
Univ., Ser. Math.and Mech. 54, 2 (1999).}

\bibitem{BEG} {\footnotesize A.Braverman, P.Etingof, D.Gaitsgory "Quantum
integrable systems and differential Galois theory", Transform.Groups. 2, 1,
31-56 (1997).}

\bibitem{CMR} {\footnotesize F.Calogero, C.Marchioro, O.Ragnisco "Exact
solution of the classical and quantal one-dimensional many-body problems
with the two-body potential }$V_{a}(x)=g^{2}a^{2}/sh^{2}ax${\footnotesize ",
Lett.Nuovo Cim.13(10), 383--387 (1975).}

\bibitem{B} {\footnotesize F.A.Berezin "Laplace operators on semisimple Lie
groups", Proc. Moscow Math. Soc.6, 371--463 (1957).}

\bibitem{WW} {\footnotesize E.T.Whittaker, G.N.Watson "A course of modern
analysis", Cambridge University Press, (1963).}
\end{thebibliography}
\end{document}